\documentstyle[12pt]{article}

\newcommand{\be}{\begin{equation}}
\newcommand{\ee}{\end{equation}}
\newcommand{\bea}{\begin{eqnarray}}
\newcommand{\eea}{\end{eqnarray}}
\newcommand{\nn}{\nonumber \\}

\newcommand{\ba}{\begin{array}}
\newcommand{\ea}{\end{array}}
\newcommand{\vs}[1]{\vspace{#1 mm}}

\def\ie{{\it i.e.,}\ }

\def\bbox{{\,\lower0.9pt\vbox{\hrule \hbox{\vrule height 0.2 cm
\hskip 0.2 cm \vrule height 0.2 cm}\hrule}\,}}
\newcommand{\dsl}{\pa \kern-0.5em /}
\newcommand{\la}{\lambda}
\newcommand{\veps}{\varepsilon}

\newcommand{\pa}{\partial}

\font\mybb=msbm10 at 12pt
\def\bb#1{\hbox{\mybb#1}}

\def\bR {\bb{R}}
\def\bE {\bb{E}}

\def\bC {\bb{C}}

\newcommand{\reef}[1]{(\ref{#1})}

\def\l{\lambda}

\def\cosec{\,{\rm cosec}\,}

\begin{document}

\topmargin 0pt
\oddsidemargin 5mm

\renewcommand{\thefootnote}{\fnsymbol{footnote}}
\begin{titlepage}

\setcounter{page}{0}

\rightline{\small hep-th/9810204  \hfill QMW-PH-98-38}
\vs{-3}
\rightline{\small \hfill DAMTP-1998-132}
\vs{-3}
\rightline{\small \hfill McGill/98-29}
\vskip 1cm

\vs{5}
\begin{center}
{\Large BLACK HOLES OF D=5 SUPERGRAVITY}
\vs{10}

{\large
Jerome P. Gauntlett$^1$, Robert C. Myers$^2$ and  Paul K. Townsend$^3$} \\
\vs{5}
${}^1${\em Department of Physics\\
       Queen Mary and Westfield College\\
       University of London, Mile End Road\\
       London E1 4NS, UK}\\
\vs{3}
${}^2${\em Department of Physics, McGill University\\
Montr{\'e}al, PQ, H3A 2T8, Canada}\\
\vs{3}
${}^3${\em DAMTP, University of Cambridge\\ 
Silver Street, Cambridge, CB3 9EW, UK}
\end{center}
\vs{7}
\centerline{{\bf Abstract}}

We discuss some general features of black holes of five-dimensional
supergravity, such as the first law of black hole mechanics. We also discuss 
some special features of rotating supersymmetric black holes. In particular,
we show that the horizon is a non-singular, and {\sl non-rotating}, null
hypersurface whose intersection with a Cauchy surface is a squashed 3-sphere. We
find the Killing spinors of the near-horizon geometry and thereby determine the
near-horizon isometry supergroup.

\vs{5}

\end{titlepage}
\newpage
\renewcommand{\thefootnote}{\arabic{footnote}}
\setcounter{footnote}{0}

\section{Introduction}

Stationary, charged, asymptotically-flat black hole solutions of the vacuum
Einstein-Maxwell equations exist for all spacetime dimensions $D\ge4$
\cite{Tang,MP}. They are characterized by their mass $M$, charge $Q$, and a
number of angular momenta ${\bf J}$ equal to the  rank of the rotation group
$SO(D-1)$. They have an event horizon with surface gravity $\kappa$ and
$(D-2)$-volume
${\cal A}$, electric potential $\Phi_H$ and angular velocities ${\bf \Omega}_H$.
These quantities are related by the first law of black hole
mechanics 
\be\label{first} 
dM= {\kappa\over 8\pi G_D} d{\cal A} + \Phi_H dQ + 
{\bf \Omega}_H\cdot d{\bf J}\, .
\ee
where $G_D$ is the $D$-dimensional Newton constant. This law was first
established in the context of $D=4$ Einstein-Maxwell theory \cite{BCH}. It
has been generalized to arbitrary $D$ for uncharged black holes \cite{MP,Wald}.
One purpose of this paper is to provide the further generalization to
charged black holes.

We shall be interested principally in {\sl supersymmetric} charged black holes.
These are naturally viewed as (special) solutions of $D$-dimensional
supergravity theories, but they can be defined independently of supergravity as
black hole solutions admitting at least one Killing spinor. Remarkably, such
solutions exist {\sl only} for $D=4$ and $D=5$. The $D=4$ case is already
well-understood, so the main focus of this paper will be on the
$D=5$ case. We should emphasize that by `black hole' we  mean here an
asymptotically flat spacetime that is non-singular on and outside an event
horizon. There are supersymmetric particle-like solutions of
$D>5$ supergravity theories that are sometimes called black holes, but these are
always singular. There are also supersymmetric black holes in
$D=3$, but the spacetime in that case is asymptotically anti-de Sitter rather
than asymptotically flat. Of course, there are non-singular supersymmetric black
brane solutions in various $D\ge4$ supergravity theories but these are neither
`particle-like' nor, strictly speaking, asymptotically flat. 

The bosonic sector of $N=2$ $D=4$ supergravity is Einstein-Maxwell theory. 
{}For any asymptotically flat solution of Einstein-Maxwell theory that is
non-singular on and outside the event horizon the mass $M$ is bounded below by
the charge $Q$ \cite{GH}. In geometrized units of charge the bound is $M\ge
|Q|$. A quantum version of this bound follows directly from the supersymmetry
algebra \cite{WO}, and configurations that saturate it, i.e. those with $M=|Q|$,
preserve half the supersymmetry of the vacuum. These configurations are the
extreme Reissner-Nordstr{\"o}m (RN) (multi) black holes, and the asymptotic
values of the Killing spinor fields that they admit can be identified with the
zero-eigenvalue eigenspinors of the anticommutator of supersymmetry charges.
Although the bound $M\ge |Q|$ makes no mention of the angular momentum 
$J$, solutions that saturate it are singular\footnote{The singularity
may be resolved in Kaluza-Klein theory \cite{Tada}.} unless
$J=0$. Another way to understand why the angular momentum must vanish for
supersymmetric $D=4$ black holes is to observe that the unique Killing vector
field $k$ with normalization
$k^2=-1$ at infinity is the `square' of a non-vanishing Killing spinor that is
non-singular everywhere outside the horizon. This implies that $k$ is timelike
everywhere outside the horizon and hence that there is no ergoregion, but a
black hole for which the horizon has a non-vanishing angular velocity
necessarily has an ergoregion. We thereby deduce that a supersymmetric black
hole must be `non-rotating' in the sense that its horizon cannot rotate. It then
follows that $J=0$,
e.g. from the theorem that {\it every stationary, non-rotating Einstein-Maxwell
black hole must be static} \cite{Waldb}. 

When considering how these results about $D=4$ Einstein-Maxwell theory may
generalize to $D=5$ it is important to appreciate that in $D=5$ it is possible
to include an additional `$AFF$' Chern-Simons (CS) term. This makes no
difference to the class of static solutions but it does affect the class of
stationary solutions. Here we shall take the $D=5$ `Einstein-Maxwell' theory to
be the bosonic sector of $D=5$ supergravity for which the CS term is present
with a particular coefficient. Specifically, the Lagrangian is \cite{Cremmer}
\be\label{crem}
L= {1\over 16\pi G_5} \left[\sqrt{-g}(R -F^2) -
{2\over 3\sqrt{3}}\varepsilon^{mnpqr}A_mF_{np}F_{qr}\right]
\ee
The mass $M$ of any asymptotically flat solution satisfies the bound
\cite{GKLTT}
\be\label{bound}
M\ge {\sqrt{3}\over 2}|Q|
\ee
and solutions that saturate this bound admit Killing spinors. 

The general stationary black hole solution of $D=5$ supergravity will depend on
four parameters, the mass $M$, the electric charge $Q$, and two angular momenta,
$J_1$ and $J_2$. It is implicit in the 7-parameter string solution of a $D=6$
model discussed in \cite{lessix}, since the $D=5$ Lagrangian (\ref{crem}) is a
truncation of an $S^1$ compactification of the $D=6$ Lagrangian used there, but
we shall not need the explicit solution of \cite{lessix} here. The bound
(\ref{bound}) makes no reference to the two angular momenta $J_1$ and $J_2$ but,
as in the $D=4$ case, the requirement of non-singularity constrains these
parameters.  When the bound is saturated the constraint is such that one linear
combination  of $J_1$ and $J_2$ must vanish. Thus, the  general {\sl
supersymmetric} $D=5$ black hole \cite{tsey,myers} (see also \cite{sabra}) is
parameterized  by its mass $M$ and one angular  momentum, which we shall call
$J$. The $J=0$ case is just a straightforward generalization to $D=5$ of the
extreme RN solution, and it is sometimes called the `Tangherlini' black hole
\cite{Tang}; in the context of
$D=5$ supergravity  it preserves half the supersymmetry of the vacuum
\cite{GKLTT}. 

The fact that there exist {\sl rotating} supersymmetric black holes is quite
surprising in view of the previous argument that the horizon must be
non-rotating. That argument was presented in the context of $D=4$ black holes
but it is equally valid for $D=5$. Thus, it is {\sl not} true that every
stationary, non-rotating $D=5$ Einstein-Maxwell black hole is static, if by
`non-rotating' one means vanishing angular velocity of the horizon, and by
`Einstein-Maxwell' one means the bosonic sector of $D=5$ supergravity. The
supersymmetric rotating black holes are counter-examples to this would-be
theorem\footnote{An apparent further counter-example in $D\ge6$ 
is provided by a
class of rotating black holes for which the angular velocity of the
horizon, but not the total angular momentum, vanishes in an extremal 
limit \cite{HSen}. But the $(D-2)$-volume of the horizon of these solutions
vanishes in the same limit, so a non-zero angular velocity would have been
paradoxical.}. We should stress that these are solutions of the
`Einstein-Maxwell' equations only if the latter are understood to include the
contribution of the CS term with the precise coefficient required by $D=5$
supergravity. It is quite possible that this particular theory is exceptional in
this regard among the class of Einstein-Maxwell theories parameterized by the CS
coefficient. 

These observations indicate that a thorough investigation of the properties
of black hole solutions of $D=5$ supergravity is warranted, supersymmetric black
holes in particular. Here we begin this investigation with a derivation of the
first law of black hole mechanics (generalizing the result of \cite{MP} for
uncharged black holes) and a study of the local and global properties of
supersymmetric black holes. In particular, we confirm that a rotating
supersymmetric black hole has a non-singular and non-rotating horizon. The
effect of rotation on the horizon is not to make it rotate but to deform it from
a round 3-sphere to a squashed 3-sphere! This implies that the angular
momentum is stored in the Maxwell field, which it is, but another surprise is
that a negative fraction of the total is stored {\sl behind} the horizon.

It has been appreciated for some time that the extreme RN black hole 
interpolates between D=4 Minkowski spacetime (near infinity) and the
Robinson-Bertotti $adS_2\times S^2$ spacetime (near the horizon) \cite{Carter}
and that the latter is, like D=4 Minkowski, a maximally supersymmetric vacuum
solution of N=2 D=4 supergravity \cite{Gibbons}. In fact, it admits an
$SU(1,1|2)$ isometry supergroup \cite{Kal}. There is therefore a restoration of
full supersymmetry in either of the two asymptotic regions. These results have a
direct generalization to the Tangherlini black hole. This solution interpolates
between maximally supersymmetric vacua of $D=5$ supergravity, in this case
between $D=5$ Minkowski spacetime (near spatial infinity) and $adS_2\times S^3$
(near the horizon) \cite{GKLTT}. There is again a full restoration of
supersymmetry in either asymptotic region \cite{Ferrara}. Here we are able to
identify the isometry supergroup in the near-horizon limit as $SU(1,1|2)\times
SU(2)_R$ where $SU(2)_R$ is the group generated  by the left-invariant vector
fields on $S^3\cong SU(2)$ and $SU(1,1|2)$  contains the $SU(2)_L$ subgroup
generated by the right-invariant vector fields. Apart from the $SU(2)_R$ factor
this is the same as the D=4 case\footnote{This might seem surprising but it is a
reflection of the fact that the $SU(2)$ gauge fields in the effective D=2
supergravity theory arising from compactification of $D=5$ supergravity on $S^3$
belong not to the graviton supermultiplet but to an $SU(2)$ super-Yang-Mills
multiplet.}.

Like the (non-rotating) Tangherlini black hole, the rotating supersymmetric 
$D=5$ black hole also preserves half the supersymmetry of the vacuum. Moreover,
it was shown in \cite{Renata} that there is again a full restoration of
supersymmetry near the horizon. We confirm this result here by a direct
determination of the Killing spinors. We are then able to show, using a method
introduced in \cite{GMT}, that the isometry supergroup of the near-horizon
limit is
\be
SU(1,1|2)\times U(1)_R\, .
\ee
In other words, the rotation breaks $SU(2)_R$ to $U(1)_R$ without affecting
the supersymmetries. The near-horizon spacetime is a
homogeneous spacetime of the form
\be
[SO(2,1)\times SU(2)_L \times U(1)_R]/[U(1)\times U(1)]\, .
\ee
{\sl One} such space is the direct product of $adS_2$ with a squashed 3-sphere. 
The horizon (or rather, its intersection with a Cauchy surface) is indeed a
squashed 3-sphere, with a squashing parameter simply related to $J$, but the 
full five-dimensional spacetime is {\sl not} a direct product. 

The organization of this paper is as follows. We begin with a derivation of the
first law of black hole mechanics for black holes of $D=5$ supergravity. We
then solve the Killing spinor equations to recover the explicit rotating
supersymmetric black hole solutions of \cite{myers}. We then determine various
properties of the horizon, of the global structure behind the horizon, and we
study the distribution of the angular momentum in the Maxwell field. 
We follow this with a detailed analysis of the near-horizon limit, its Killing
spinors and its isometry supergroup. We conclude with some speculations about
how the physics of $D=5$ black holes in general Einstein-Maxwell-CS theories
might depend on the CS coefficient.

\section{The First Law}

The bosonic fields of minimal D=5 supergravity are the 5-metric $g$ and a 
Maxwell 1-form $A$ with 2-form field strength $F=dA$. The field equations
include Einstein's equations (for coordinates $x^m$, $m=0,1,\dots,4$)
\be\label{fieldeq}
G_{mn} = 2T_{mn}
\ee
with the Maxwell stress-energy tensor
\be\label{stress}
T_{mn} = F_{mp}F_n{}^p - {1\over4}g_{mn} F^2\ ,
\ee
and Maxwell's equations modified by the CS contribution
\be\label{fieldeqb}
D_mF^{mn} = {1\over 2\sqrt{3}\sqrt{-g}}
\varepsilon^{npqrs}F_{pq}F_{rs}\ . 
\ee
Note that the stress-energy tensor is covariantly conserved (i.e.
$D_mT^{mn}=0$) for solutions of the Maxwell/CS field equations by virtue of
the five-dimensional identity
\be\label{ident}
{}F_{[mn}F_{pq}F_{r]s} \equiv 0\, .
\ee

Our main aim in this section will be to derive a Smarr-type formula for
(stationary and asymptotically flat) black hole solutions of the above
equations. We will then use this to derive the first law \reef{first}. 
In attacking this problem, however, it will be useful to consider first the
pure Einstein-Maxwell theory in which (\ref{fieldeqb}) is replaced by the
simpler Maxwell-equation
\be\label{maxwell}
D_mF^{mn}=0\ .
\ee
In this case the extension to arbitrary spacetime dimension $D$ is
straightforward, and will facilitate the comparison of the $D=4$ and $D=5$ 
cases. We will then extend the analysis to the case of odd dimensions
$D=2n+1$ in which the Einstein-Maxwell action may be supplemented by an
`$A\,F^n$' CS term. This will include the $D=5$ case of interest, which will
turn out to be a rather special case among the class of odd-dimensional
Einstein-Maxwell-CS theories. 

Let us begin with the general analysis of the Einstein-Maxwell system.
We shall be concerned with stationary asymptotically flat solutions, for which
there exists a unique (timelike) Killing vector field $k$ normalized such that
$k^2=-1$ at spatial infinity. 
{}For $D$ spacetime dimensions (with $D\ge4$) the total mass is given by
\cite{MP}
\be\label{komar}
M = -{(D-2)\over (D-3) 16\pi G_D} \oint_\infty dS_{mn} D^mk^n\, ,
\ee
where the integral is taken over the $(D-2)$-sphere at spatial infinity.

We shall further restrict our attention to solutions admitting an additional
$[(D-1)/2]$ commuting spacelike vector fields $\bf m$ with closed
orbits; these are associated with the angular momenta ${\bf J}$ \cite{MP}; there
will be two such Killing vector fields for $D=5$ corresponding to the two
angular momenta $J_1$ and $J_2$. We can choose coordinates such that
$m_i=\partial/\partial\varphi^i$ and since the orbits are closed we can
normalize ${\bf m}$ by requiring the coordinates $\varphi^i$ to be identified
modulo $2\pi$. The associated angular momenta are then \cite{MP}
\be\label{angular}
{\bf J}= {1\over 16\pi G_D} \oint_\infty dS_{mn}D^m{\bf m}^n\, .
\ee
{}Finally, the total electric charge, in geometrized units, is
\be
Q = {1\over 8\pi G_D}\oint_\infty dS_{mn}F^{mn}\, .
\ee

Our derivation of the first law relies on the theorem that the event
horizon of a stationary black hole is a Killing horizon of some linear
combination of $k$ and ${\bf m}$ \cite{Hawking}. Let
\be\label{hawk}
\xi = k+ {\bf \Omega}_H\cdot {\bf m}
\ee
be this Killing vector field. The constant coefficients ${\bf \Omega}_H$ are the
angular velocities of the horizon. Let $\Sigma$ be a spacelike hypersurface
with boundaries at spatial infinity and on the horizon. Using Gauss' law and
(\ref{hawk}), we can then rewrite each of the formulae for
$M$ and ${\bf J}$ as the sum of an integral over $\Sigma$ and a surface
integral over the boundary $H$ of $\Sigma$ on the horizon. This leads to the
formula
\bea\label{MandJ}
M &=& -{(D-2)\over (D-3)4\pi G_D}\int_\Sigma dS_m R^m{}_n\xi^n \nn 
&& + {(D-2)\over (D-3)}{\bf \Omega}_H\cdot {\bf J} 
-{(D-2)\over (D-3)16\pi G_D}\oint_H dS_{mn} D^m\xi^n 
\eea

We may write $dS_{mn} = 2d{\cal A}\, \xi_{[m}n_{n]}$ for some null vector
field $n$ such that $\xi\cdot n=-1$. Then, using the fact that $(\xi\cdot D)\xi
=\kappa\xi$ on the horizon, where $\kappa$ is the horizon surface gravity,
which is constant by the zeroth law, we  may express the final integral in terms
of $\kappa$  and the $(D-2)$-volume of the horizon ${\cal A}$.  If in addition
we use the Einstein equation in the form
\be
R^m{}_n = F^{mp}F_{np} - {1\over 2(D-2)}\delta^m_n F^2\, ,
\ee
then we arrive at the formula
\bea\label{Minter}
M &=& -{(D-2)\over (D-3)4\pi G_D}\int_\Sigma dS_m\big[F^{mp}(\xi^n F_{np}) -
{1\over 2(D-2)}\xi^m F^2\big] \nn
&& \qquad + {(D-2)\over (D-3)}{\bf \Omega}_H\cdot {\bf J}
+ {(D-2)\over (D-3)8\pi G_D}\, \kappa {\cal A}
\eea

We shall assume that ${\cal L}_\xi F=0$, where ${\cal L}_\xi$ is the Lie 
derivative with respect to the vector field $\xi$. By a choice of gauge we can
then arrange that ${\cal L}_\xi A =0$, from which it follows that
\be\label{gaugecon}
\xi^nF_{np} = -\partial_p(\xi\cdot A)\, ,
\ee
and hence that
\be\label{onea}
{}F^{mp}(\xi^n F_{np})= -D_p\left[(\xi\cdot A) F^{mp}\right] -(\xi\cdot
A)(D_pF^{pm})\, .
\ee
It also follows that
\be\label{oneb}
\xi^mF^2 = 4D_p\left[\xi^{[m}F^{p]n}A_n\right] - 2\xi^mA_n(D_pF^{pn})\, .
\ee
Using these relations we find that
\bea\label{twoa}
M &=& {(D-2)\over (D-3)4\pi G_D} \int_\Sigma 
dS_mD_p\left[(\xi\cdot A) F^{mp}
+ {2\over (D-2)}\xi^{[m}F^{p]n}A_n\right] \nn
&& +{1\over (D-3)4\pi G_D}\int_\Sigma dS_m\left[
(D-2)(\xi\cdot A)\delta^m_n-\xi^mA_n\right](D_pF^{pn}) \nn
&& + {(D-2)\over (D-3)}{\bf \Omega}_H\cdot {\bf J} + {(D-2)\over (D-3)8\pi
G_D}\,\kappa {\cal A} 
\eea
The second of the above integrals vanishes for solutions of the Maxwell
equations \reef{maxwell}; we keep it here because it will not vanish
when CS interactions are added. The first integral can be rewritten as the
difference of surface integrals at spatial infinity and on the horizon. We shall
suppose that $A$ falls off sufficiently fast near spatial infinity that the
surface integral at spatial infinity vanishes. We are then left with the surface
integral
\be\label{expr}
-{(D-2)\over (D-3)8\pi G_D} \oint_H dS_{mp}\left[(\xi\cdot A) F^{mp} +
{2\over(D-2)}\xi^{m}F^{pn}A_n\right]\, .
\ee

Now, Raychaudhuri's equation for geodesic deviation implies that the scalar
$R_{mn}\xi^m\xi^n$ vanishes on a Killing horizon of $\xi$. Using the Einstein
equation and the fact that $\xi$ is null on the horizon we deduce 
that the 1-form $V=i_\xi F$ is also null on the horizon. But, $\xi\cdot V\equiv
0$ so $V^m\partial_m$ is tangent to the horizon. Any tangent to the horizon
that is null must be proportional to $\xi$, so we deduce that, on the horizon,
$\xi^mF_{mn}$ is some function times $ \xi_n$. It then follows (by the use of the
relation $dS_{mn} = 2d{\cal A}\,\xi_{[m}n_{n]}$) that
\be
\oint_H dS_{mp}\xi^{m}F^{pn}A_n = -{1\over2}
\oint_H dS_{mp} (\xi\cdot A) F^{mp}\, ,
\ee
which allows us to reduce (\ref{expr}) to
\be\label{expb}
- {1\over 8\pi G_D}\oint_H dS_{mn} (\xi\cdot A) F^{mn}
\ee
Now $(\xi\cdot A)$ is constant on the horizon\footnote{From
\reef{gaugecon} we have $d(\xi\cdot A)=-i_\xi F$. As explained above, $i_\xi F
\propto dx^m\xi_m$ on the horizon. It follows that $i_t d(\xi\cdot A)_H \propto
t\cdot \xi$, which vanishes if $t$ is tangent to the horizon.}. The constant
is, by definition, minus the co-rotating electric potential
$\Phi_H$, i.e. $(\xi\cdot A)_H=-\Phi_H$. Taking this constant outside the
integral, an application of Gauss' theorem then yields
\be
- {1\over 8\pi G_D}\oint_H dS_{mn} (\xi\cdot A) F^{mn} =
\Phi_H \left[ Q +  {1\over 4\pi G_D}\int_\Sigma dS_m D_pF^{pm}\right]\, .
\ee
If we now assume the validity of the Maxwell equations \reef{maxwell} then we 
may rewrite the expression (\ref{twoa}) for $M$ as
\be\label{smarr}
M= \Phi_H Q + {(D-2)\over (D-3)}{\bf \Omega}_H\cdot {\bf J} + 
{(D-2)\over (D-3)8\pi G_D}\, \kappa {\cal A}\ .
\ee

We have thus arrived at a Smarr-type formula for $D$-dimensional Einstein-Maxwell
black holes from which the first law can be deduced. We shall begin by assuming
that black hole solutions are uniquely determined by their mass, charge, and
angular momenta. This is what one would expect if the uniqueness theorems
proved for $D=4$ can be extended to $D=5$. In this case the mass is a function
$M(Q,{\bf J},{\cal A})$ of the charge, angular momenta, and $(D-2)$-volume of
the horizon cross-section. In units for which $G_D=1$ the independent variables
of this function have dimensions as follows:
\be
[Q] =M\, ,\qquad [J] = [{\cal A}] =M^{(D-2)/(D-3)}\, .
\ee 
There can be no dependence of the function $M$ on any other dimensionful
quantity because the Einstein-Maxwell Lagrangian involves no dimensional
constants other than the overall factor of $1/16\pi G_D$. This is because every
term in the Lagrangian has precisely two derivatives (and we take the metric and
gauge potential to be dimensionless). It follows that $M(Q,{\bf J},{\cal A})$
must be a weighted homogeneous function, to which an application of Euler's
theorem yields
\be\label{partials}
Q{\partial M\over\partial Q} + {(D-2)\over (D-3)} 
{\bf J}\cdot{\partial M\over\partial {\bf J}} 
+ {(D-2)\over (D-3)} {\cal A}{\partial M\over\partial {\cal A}}
= M
\ee
Substituting (\ref{smarr}) on the right hand side, and using the independence of
the variables ($Q,{\bf J},{\cal A}$), we deduce (reinstating the
$G_D$-dependence) that
\be
{\partial M\over\partial Q} = \Phi_H \quad, \qquad
{\partial M\over\partial {\bf J}} = {\bf \Omega}_H\quad,\qquad
{\partial M\over\partial {\cal A}} = {\kappa\over 8\pi G_D}
\ee
and hence the first law (\ref{first}). 

Of course, the preceding analysis does not apply to the five-dimensional
theory in which we were originally interested because of the CS
term in (\ref{crem}). More generally, it fails to apply in any odd spacetime
dimension $D=2n+1$ for which the Lagrangian includes a CS term, which
we may write as 
\be\label{csact}
L_{\rm int}=-{1\over16\pi G_D}{4\la\over n+1}\veps^{mnp\cdots qr}
A_mF_{np}\cdots F_{qr}
\ee
for some coupling constant $\la$. With the addition of this term the equation 
of motion for $A$ becomes
\be\label{csmot}
D_mF^{mn} = {\la\over\sqrt{-g}}
\varepsilon^{npq\cdots rs}F_{pq}\cdots F_{rs}\ . 
\ee
It will be useful to rewrite this equation as
\be\label{csmotb}
D_m\left(F^{mn}+2\la J^{mn}\right)=0
\ee
where
\be\label{extra}
J^{mn}={1\over\sqrt{-g}}\veps^{mnpqr\cdots st}A_pF_{qr}\cdots F_{st}\ .
\ee
We also note that the five-dimensional identity \reef{ident}, which
ensured stress-energy conservation, is extended to
\be\label{identb}
{}F_{[mn}F_{pq}\cdots F_{r]s} \equiv 0
\ee
with $n+1$ factors of the field strength in $D=2n+1$ dimensions.

Most of the previous analysis remains unchanged, but we must now reconsider the
integral
\be\label{intx}
{1\over (D-3)4\pi G_D}\int_\Sigma dS_m\left[
(D-2)(\xi\cdot A)\delta^m_n-\xi^mA_n\right](D_pF^{pn})
\ee
in (\ref{twoa}). This contribution was previously zero. Now, as a result of
(\ref{csmotb}), it equals
\be
-{\la\over (D-3)2\pi G_D}\int_\Sigma dS_m\left[
(D-2)(\xi\cdot A)\delta^m_n-\xi^mA_n\right](D_pJ^{pn})
\ee
Using the identity $A_nJ^{np}\equiv 0$, we may simplify this to
\be\label{intxa}
-{\la\over (D-3)4\pi G_D}\int_\Sigma dS_m\left[
2(D-2)(\xi\cdot A)D_pJ^{pm} + \xi^mF_{pn}J^{pn}\right]
\ee
Now, the identity $\xi^{[m}\varepsilon^{npq\dots rs]}\equiv 0$ can be used to
show that
\be\label{wow}
\xi^m F_{pn}J^{pn}= -2 (\xi\cdot A) D_p J^{pm} - (D-1)\partial_n(\xi\cdot
A)J^{mn}
\ee
which allows us to rewrite the expression (\ref{intxa}) as
\be
-{\la\over (D-3)2\pi G_D}\int_\Sigma dS_m\left[2(D-3)D_p(\xi\cdot A\, J^{pm})
-(D-5)\partial_p(\xi\cdot A) J^{pm}\right]
\ee
Under the same assumptions as before regarding the behaviour of $A$ near
spatial infinity, and again using (\ref{gaugecon}), this equals
\be
- {\la\over 4\pi G_D}\oint_H dS_{mp}(\xi\cdot A) J^{mp} -
{(D-5)\la\over (D-3)4\pi G_D}\int_\Sigma dS_m\,
\xi^nF_{np}J^{pm}
\ee
The surface term may be combined with that in (\ref{expb}) to yield
\be\label{expbb}
- {1\over 8\pi G_D}\oint_H dS_{mn} (\xi\cdot A)\left[ F^{mn} +2\la
J^{mn}\right]
\ee
As before $(\xi\cdot A)$ is constant on the horizon and can be taken outside the
integral. Using Gauss' theorem and the modified equation of motion \reef{csmotb}
we may reduce this surface term to $\Phi_H\,Q$, so our expression
for $M$ becomes
\be\label{smarrb}
M= \Phi_H Q + {(D-2)\over (D-3)}{\bf \Omega}_H\cdot {\bf J} + 
{(D-2)\over (D-3)8\pi G_D}\, \kappa {\cal A}+{(D-5)\over(D-3)}\la\,I\ .
\ee
where 
\be\label{leftover}
I=-{1\over 4\pi G_D}\int_\Sigma dS_m
\xi^nF_{np}J^{pm}\ .
\ee

The case of principal interest here is $D=5$ for which the coefficient of
the last term in (\ref{smarrb}) vanishes and we
therefore recover the Smarr-type formula used previously to derive the first
law. The $D=5$ case is also special in one other regard: the CS term is
quadratic in derivatives and therefore has the same dimension as the other
terms in the Lagrangian, which again involves no dimensionful parameters other
than the overall factor of $1/16\pi G_D$. The previous proof of the first law
therefore goes through without change. In the case of supersymmetric black
holes for which (as we shall confirm in the following section) $\kappa$ and
${\bf \Omega}_H$ vanish, we have $M=\Phi_H Q$. Because supersymmetric solutions
saturate the bound (\ref{bound}) we deduce that $\Phi_H = \sqrt{3}/2$. We shall
shortly confirm this prediction. 
 
{}For all odd spacetime dimensions higher than five, the coefficient
of the $\la\,I$ term in (\ref{smarrb}) does not vanish, with the consequence
that this formula is no longer equivalent to
(\ref{smarr}). This is not unexpected because the proof of the first law via
Euler's theorem also requires modification. This is because the CS term is now
more than quadratic in derivatives and the coupling 
constant $\lambda$ is therefore a new
dimensionful parameter. Specifically, the dimension of $\lambda$, for $G_D=1$, 
is $[\la]=(D-5)/(D-3)$. Euler's theorem is still applicable provided
that $M$ is now considered to be a function of $\lambda$ (as well as 
$Q$, ${\bf J}$ and ${\cal A})$, but this leads to the
addition of $(D-5)/(D-3)\,\la \,(\partial M/ \partial \la)$ to the left
hand side of (\ref{partials}). Comparing this new equation with (\ref{smarrb})
we deduce that
\be
I= {\partial M\over\partial\lambda} 
\ee
and hence that
\be
dM= {\kappa\over 8\pi G_D} d{\cal A} + \Phi_H dQ + 
{\bf \Omega}_H\cdot d{\bf J} + Id\lambda\, .
\ee
This is a version of the first law that is reminiscent of its extension in
\cite{GKK} to include scalar vacuum expectation values. In our case,
however, the parameter $\lambda$ labels {\sl theories} rather than solutions, so
for any given theory $\lambda$ is fixed, the last term vanishes, and we
recover the first law as it appears in (\ref{first}). We thus see, not
unexpectedly, that the first law is quite robust and does not depend
on special features of D=5 supergravity. Nevertheless, the black holes of D=5
supergravity do have many special features, and we shall explore some of them 
in the remainder of this article. 

\section{Supersymmetric black holes}

Supersymmetric solutions of the field equations (\ref{fieldeq}) are those for
which there exist non-vanishing solutions for $\zeta$ of the Killing spinor
equation \cite{GKLTT}
\be\label{kilspin}
\big[d + {1\over4}\omega_{ab}\Gamma^{ab} + {i\over 4\sqrt{3}}
\left(e^a\Gamma^{bc}{}_aF_{bc} - 4e^a\Gamma^bF_{ab}\right) \big]\zeta
=0
\ee
where $e^a$ are the frame 1-forms and $\omega_{ab}$ is the spin-connection 
1-form\footnote{In contrast to \cite{GKLTT} we use here the `mostly plus' metric
convention, which accounts for the factor of $i$.}. The spinor
$\zeta$ is necessarily complex. The Dirac matrices are also complex but the
product of all five is $\pm i$. A choice of the sign amounts to a choice of one
of two inequivalent $4\times 4$ representations. We shall choose the Dirac
matrices such that
\be\label{inequiv}
\Gamma^{01234}=i\, .
\ee

We shall consider configurations of the form
\be\label{fivesol}
ds^2 = -(e^0)^2 + e^ie^j\delta_{ij}\quad,\qquad
{}F = {\sqrt{3}\over2} de^0\quad,
\ee
where
\be\label{framed}
e^0 = H^{-1}(dt+a) \quad, \qquad e^i = H^{1\over2} dx^i \qquad (i=1,2,3,4).
\ee
with time-independent function $H$ and 1-form $a$. A calculation yields
\bea
\omega_{0i} &=& e^0 H^{-3/2}\partial_i H - {1\over2} e^j H^{-2} f_{ij}\\
\omega_{ij} &=& e^k H^{-3/2}\delta_{k[i}\partial_{j]}H +
{1\over2}e^0H^{-2}f_{ij} \\
{}F &=& -{\sqrt{3}\over2}H^{-3/2}\partial_iH\, e^i\wedge e^0 + {\sqrt{3}\over4}
H^{-2}f_{ij}\, e^i\wedge e^j
\eea
where
\be
f_{ij} = \partial_i a_j -\partial_j a_i\, .
\ee
Since
\be 
d= e^0 H\partial_t + e^i H^{-1/2}(\partial_i - a_i \partial_t)
\ee
we find that the Killing spinor equations become
\bea
\partial_t\zeta &=& {1\over2}H^{-3}\left(iH^{1/2}\partial_iH\Gamma^i -
{1\over4} f_{ij}\Gamma^{ij}\right)(1-i\Gamma^0)\zeta \nn
\left(\partial_i -a_i\partial_t\right)\zeta &=& -{1\over4}H^{-3/2}\partial_j H
\Gamma^{ij}(1-i\Gamma^0)\zeta + i\partial_i\left(\log
H^{-1/2}\right)\Gamma^0\zeta \nn
&&\qquad + {1\over4}\left(f_{ij}+\tilde f_{ij}\right)\Gamma^j\Gamma^0\zeta
\eea 
where
\be
\tilde f_{ij} = {1\over2}\epsilon_{ijkl}f_{kl}\, .
\ee
The integrability conditions are consistent with non-zero $\zeta$ only if
\be
i\Gamma^0\zeta =\zeta\, ,
\ee
and
\be\label{selfd}
f_{ij}+\tilde f_{ij}=0\, .
\ee
The Killing spinor equations are then solved by
\be
\zeta = H^{-1/2}\zeta_0
\ee
where $\zeta_0$ is a constant eigenspinor of $i\Gamma^0$ with eigenvalue $1$.
The Maxwell field equation implies that $H$ is harmonic. Allowing for point
singularities, we have
\be
H= 1 +\sum_i \mu_i/|x-x_i|^2
\ee
where $x$ is a displacement vector in $\bE^4$ and $\mu_i$ are a
set of positive constants. 

The anti-self-duality equation (\ref{selfd}) for $f=da$ is solved by 
\be
a= dx^m J_m{}^n\partial_n K
\ee
where $K$ is a harmonic function on $\bC^2\cong \bE^4$ with complex structure
$J$. If we require that $a$ vanish at spatial infinity then any non-constant
$K$ must have singularities, but these can be chosen to be point
singularities that coincide with the singularities of $H$. Thus each
singularity of $H$ is associated with some residue of the function $K$.  
Under circumstances to be spelled out below these singularities are merely
coordinate singularities at Killing horizons of the metric, the union of which
is the event horizon. For $A$ to vanish at infinity as assumed in our
discussion of the first law we must choose the Maxwell gauge such that
\be
A= {\sqrt{3}\over2} \left[ H^{-1}(dt +a) -dt\right]\, .
\ee
Since $H^{-1}$ vanishes on the event horizon we deduce that $\Phi_H$ takes the
same constant value, $\sqrt{3}/2$, on each connected component of the horizon. 
As pointed out in our earlier discussion this value is required  by any black
hole solution that saturates the bound (\ref{bound}). 

If we now assume a connected event horizon then the harmonic function $H$ has a
single point singularity  in $\bE^4$, which we may choose to be the origin of
spherical polar coordinates. Thus, the $\bE^4$ metric is
\be
ds^2(\bE^4) = dr^2 + r^2 d\Omega_3^2
\ee
where $d\Omega_3^2$ is the $SO(4)$ invariant metric on the 3-sphere. 
In these coordinates we have
\be
H= 1 + {\mu\over r^2}
\ee
where $\mu$ is a positive constant related to the mass $M$ by
\be
M= {3\pi\mu \over 4G_5}\quad.
\ee
We may write the 3-sphere metric as
\be
d\Omega_3 = {1\over4}\big(\sigma_1^2 + \sigma_2^2 + \sigma_3^2\big)
\ee
where $\sigma_i$ ($i=1,2,3$) are the three left-invariant one-forms satisfying
\be
d\sigma_1 = \sigma_2 \wedge \sigma_3 \qquad {\rm and \ cyclic}
\ee
A useful explicit choice of coordinates on $S^3$ is
\be
d\Omega_3^2 = {1\over4} [d\theta^2 + d\phi^2 + d\psi^2 + 
2\cos\theta d\psi d\phi]\, ,
\ee
where
\be
0\le \theta < \pi\, ,\qquad 0\le \phi < 2\pi\, ,\qquad 
0\le \psi < 4\pi\, .
\ee
With this choice, the three one-forms $\sigma_i$ may be written as
\bea
\sigma_1 &=& -\sin\psi\, d\theta + \cos\psi\sin\theta\, d\phi \nn
\sigma_2 &=& \cos\psi\, d\theta + \sin\psi\sin\theta\, d\phi \nn
\sigma_3 &=& d\psi +\cos\theta\, d\phi
\eea

If the vector fields $\partial/\partial\psi$ and $\partial/\partial\phi$  are
assumed to be Killing then the 1-form $a$ can be assumed, without loss of
generality, to be
\be
a= {j\over 2r^2}\sigma_3
\ee
where $j$ is a parameter related to the angular momentum $J$ associated with
the Killing vector field $2\partial/\partial\psi$ (which has the normalization
assumed earlier)\footnote{Note that in the `Cartesian' coordinates of
eqs.~(\ref{fivesol},\ref{framed}), this angular momentum is related to
simultaneous rotations in two orthogonal planes.}. Specifically,
\be\label{angmom}
J= -{j\pi\over 2 G_5}\, .
 \ee
The angular momentum associated with $\partial_\phi$ vanishes. Thus, as stated
earlier, supersymmetry imposes one constraint on the two angular momenta of the
general black hole solution. 

Having specified both $H$ and $a$, we now have the explicit metric
\be
ds^2 = -\left(1+ {\mu\over r^2}\right)^{-2}
\left(dt + {j\sigma_3\over 2r^2}\right)^2
+ \left(1+ {\mu\over r^2}\right)\left(dr^2 + r^2 d\Omega_3^2\right)
\label{fivemet}
\ee
Note that $\partial_\psi$ can become timelike if
$|j|>\mu^{3/2}$, which would imply the existence of closed time-like curves
(outside the horizon). We shall therefore assume in what follows that
$|j|<\mu^{3/2}$, in which case we can write
\be
j=\mu^{3/2}\, \sin\beta
\ee
for {\sl real} $\beta$. We now turn to a study of the properties of the horizon
and the region behind the horizon.

\section{The squashed horizon and closed timelike curves}

The metric (\ref{fivemet}) is singular at $r=0$, but this is only a coordinate
singularity. To establish this we introduce the new coordinates
$\tilde t$ and $\lambda$ by
\be
t =\mu^{1/2}\tilde t \quad, \qquad  r^2= \mu(\lambda^2 -1)\quad.
\ee
The metric in these coordinates is given by $ds^2=\mu d\tilde s^2$, where
\bea\label{orig}
d\tilde s^2 &=& -\left[\left(1- {1\over \lambda^2}\right) d\tilde t{}+
{\sin\beta\over2\lambda^2} \sigma_3 \right]^2
+\left(1-{1\over \lambda^2}\right)^{-2} d\lambda^2 \nn
&& \quad+{1\over4}\lambda^2 \left(\sigma_1^2 +\sigma_2^2+\sigma_3^2\right) 
\eea
This form of the metric is essentially the same as
that originally given in \cite{myers}.

Clearly $\lambda=0,1$ are singular hypersurfaces because $g_{tt}$
blows up at $\lambda=0$ and $g_{\lambda \lambda}$ blows up at $\lambda=1$.
Everywhere else the metric is finite, and invertible provided $\theta\ne0$ since
\be
\det \tilde g = -\left({\lambda^2\over 4}\right)^3\sin^2\theta
\ee
The singularity at $\theta=0$ is the usual coordinate singularity
on the `axis' of symmetry (actually a 2-plane) of polar coordinates. The
singularity at
$\lambda=1$ (i.e. at $r=0$) may be removed as follows. We introduce new
coordinates $u,\psi'$ such that
\bea
du &=& d\tilde t - \left(1- {1\over \lambda^2}\right)^{-2} F(\lambda^2)d\lambda
\nn d\psi' &=& d\psi -2\left(1- {1\over \lambda^2}\right)^{-1}\lambda^2
G(\lambda^2)d\lambda
\label{nucord}
\eea
where $F$ and $G$ are functions of $\lambda$ to be specified later. 
We then find that
\bea
d\tilde s^2 &=&  -\left(1- {1\over \lambda^2}\right)^2 du^2 -2(F+\sin\beta\, G)du
d\lambda - \left(1- {1\over \lambda^2}\right){\sin\beta\over\lambda^2}\,
du\sigma_3' \nn
&& - (\lambda^2 -1)^{-1}\big[\sin\beta F - (\lambda^6
-\sin^2\beta)G\big]d\lambda \sigma_3' \nn
&& + {1\over4}\lambda^2(\sigma_1^2 +
\sigma_2^2) + {1\over4\lambda^4}(\lambda^6 -\sin^2\beta)(\sigma_3')^2 \nn
&& + \left(1- {1\over \lambda^2}\right)^{-2}\big[ 1-F^2 -2\sin\beta FG +
(\lambda^6 -\sin^2\beta)G^2\big] d\lambda^2 
\eea
To ensure non-singularity at $\lambda=1$ we must choose $F(\lambda)$ and
$G(\lambda)$ such that
\be
{}F(1)=\pm\cos\beta \qquad G(1)=\pm\tan\beta \qquad
F'(1)=\pm3\sin\beta\tan\beta
\ee
where the minus sign must be chosen if we wish the future horizon to be
non-singular.  A simple choice is
\be
{}F = -\left(1- {\sin^2\beta\over\lambda^6}\right)^{1/2} \qquad
G= -\sin\beta \big[\lambda^6(\lambda^6 -\sin^2\beta)\big]^{-1/2}\, ,
\ee
for which the metric is then
\bea\label{nicemetric}
\mu^{-1}d s^2 &=&  -\left(1- {1\over \lambda^2}\right)^2 du^2 + {2\lambda^3\over
(\lambda^6-\sin^2\beta)^{1/2}}\, du d\lambda - 
\left(1- {1\over \lambda^2}\right){\sin\beta\over\lambda^2}\,du\sigma_3'\nn
&& + {1\over4}\lambda^2(\sigma_1^2 + \sigma_2^2) + 
{1\over4\lambda^4}(\lambda^6 -\sin^2\beta)(\sigma_3')^2 
\eea

The limit $\lambda\rightarrow 1$ yields the near-horizon solution
\bea\label{nearhor}
d\tilde s^2 &\sim& -4\rho^2 du^2 + 2{\rm sec}\beta\, du d\rho -
2\rho\sin\beta\, du\, \sigma_3' + {1\over4}\left(\sigma_1^2 + \sigma_2^2 
+\cos^2\beta (\sigma_3')^2\right)\nn
\mu^{-1/2}F &\sim& \sqrt{3}d\rho\wedge du + {\sqrt{3}\sin\beta
\over4}\,\sigma_1\wedge \sigma_2
\eea
where $\rho=\lambda-1$. The metric is non-singular provided that
$\cos\beta\ne 0$, which we henceforth assume. Note that this solution is
invariant under the isometry
\be\label{exchange}
\rho \rightarrow -\rho \qquad u\rightarrow -u
\ee
which exchanges the region behind the Killing horizon of $k$ with the region
outside it. The maximal analytic continuation of this metric is therefore
singularity free (in contrast to the full metric) because of the absence of
singularities in the exterior region. This fact is obvious when $\sin\beta=0$
because the maximal analytic extension of the near-horizon solution is then the
direct product  of $S^3$ with the covering space of $adS_2$.

When $\sin\beta\ne0$ the intersection of the $\rho=0$ hypersurface with a 
hypersurface of constant $u$ is a squashed 3-sphere with squashing parameter
$\sin\beta$. The hypersurface $\rho=0$ is the event horizon. Its normal is
\be
\ell = g^{mn}\partial_n\rho\big|_{\rho=0} \partial_m = -\cos\beta\,
\partial_u
\ee
which is clearly null. The vector field $\partial_u$ is the timelike Killing
vector field $k$ expressed in the new coordinates. The null hypersurface
$\rho=0$ is therefore a Killing horizon of $k$ and it follows that the angular
velocity of the horizon vanishes. In other words, {\sl the rotation at infinity
corresponds to a squashed horizon rather than a rotating one}. In addition, a
calculation yields
\be
[(k\cdot D)k^m]|_{\rho=0} =0
\ee
so the event horizon has vanishing surface gravity; it is a degenerate Killing
horizon of the timelike vector field $k$. We have now verified that
supersymmetric black holes have a horizon with both vanishing surface gravity 
and vanishing angular velocities, as claimed earlier.  

The coordinates leading to (\ref{nicemetric}) have the disadvantage that there
is
a spurious coordinate singularity at $\lambda^3=\sin\beta$. To see the
significance of this coordinate singularity, and to continue the metric through
it, we instead choose
\be
{}F=\cos\beta +{3\over2}\sin\beta \tan\beta (\lambda^2-1)\qquad
G=\tan\beta
\ee
The metric is now
\bea
d\tilde s^2 &=&  -\left(1- {1\over \lambda^2}\right)^2 du^2 -2\cos\beta \big[
1+ {1\over2}\tan^2\beta (3\lambda^2-1)\big] du d\lambda
-\left(1- {1\over \lambda^2}\right){\sin\beta\over \lambda^2}\, du\sigma_3' \nn
&& + \tan\beta\big[ 1+\lambda^2 +\lambda^4 - {3\over2}\sin^2\beta\big]
d\lambda \sigma_3' + \tan^2\beta \lambda^4\left(\lambda^2 + 2
-{9\over4}\sin^2\beta\right)d\lambda^2 \nn
&& +{1\over4} \lambda^2 (\sigma_1^2 + \sigma_2^2) + 
{1\over4\lambda^4 }(\lambda^6 -\sin^2\beta)(\sigma_3')^2\, ,
\eea
which has no singularities other than the one at $\lambda=0$. That
$\lambda=0$ is a curvature singularity can be seen from the fact that
\be\label{singu}
R={2\over \l^{8}}(2\sin^2\beta-\l^2)\, .
\ee

Having passed though the horizon by one of the above two changes of coordinates
we may now change back to the original coordinates to recover the metric
(\ref{orig}) but now with the restriction to $\lambda <1$, i.e.,
\bea\label{insidemet}
d\tilde s^2 &=& -{1\over\la^4}\left[(1-\lambda^2) \, d\tilde t
+\sin\beta\sigma_3\right]^2
+ {\lambda^4 d\lambda^2\over (1-\lambda^2)^2} \nn
&& \quad+ {1\over4}\lambda^2 (\sigma_1^2 + \sigma_2^2+\sigma_3^2) 
\eea

It may appear at first sight that the global structure of this spinning
black hole coincides with that of the case $J=0$ ($\sin\beta=0$), with an
extremal
horizon surrounding a time-like and {\it point-like} singularity.
This is certainly true if one only considers radial motions, \ie
trajectories in $u$ (or $\tilde t$) and $\la$ with fixed angles.
In particular, eq.~\reef{singu}
shows that  any radial geodesic approaching $\la=0$ encounters
a curvature singularity independent of the angles. One may note though
that the angular momentum changes the nature of the singularity, in that
it increases the rate of the divergence as well as the overall sign.
Similar behavior, in particular the same angular
independence, can be seen in other curvature invariants, such
as $R_{ab}R^{ab}$
and $R_{abcd}R^{abcd}$.\footnote{This leads one to wonder what it is about the
{\sl spacetime} geometry that is `squashed'. The `squashing' appears not
to manifest itself in scalar curvature invariants, but it presumably appears
in tidal forces.} This is to be contrasted with the four-dimensional
Kerr-Newman metric, where there is a ring-like singularity which trajectories
may pass through to enter a new asymptotically flat region (with negative mass
and no horizon).

However, the simple picture above does not withstand closer scrutiny.
Firstly, the Killing vector field $m=2\partial_\psi$ becomes
null at $\lambda^3 =\sin^2\beta$ and is timelike for $\lambda^3 <\sin^2\beta$.
Since the orbits of $m$ are closed, it is clear that the region around
the singularity contains closed timelike curves. In fact, in this region, the
future light-cone tips over to encompass vectors of the form
$v(b)= -\partial_\psi+b\,\partial_u$ with $b>-(\sin\beta-\la^3)/2(1-\la^2)$ (here
we assume that $\sin\beta>0$). This can be seen to imply the existence of a
closed time-like curve passing through {\sl any} point inside the
horizon\footnote{A similar result was established in \cite{Carterctc} for the
Kerr-Newman geometry.}, as follows: from any point inside the horizon an observer
can follow a timelike trajectory into the region
$\la^3<\sin\beta$. Once there, she can continue her (non-geodesic) timelike
trajectory on an orbit of a vector field of the type
$v(b)$, travelling backward in $u$ while winding around $\psi$. Having travelled
sufficiently far back in $u$, our observer can then follow a timelike path that
moves radially outward and intersects the initial point on the trajectory. 
An alternative characterization of the hypersurface
$\lambda^3 =\sin^2\beta$ is that the normals to the hypersurfaces of constant
$t$ become null at the intersection with $\lambda^3 =\sin^2\beta$ and are
spacelike for
$\lambda^3 <\sin^2\beta$. In other words the hypersurfaces of constant $t$ are
not globally spacelike, but become timelike in the region $\lambda <
(\sin\beta)^{2/3}$. Thus one may conclude that the full
causal structure for the present spacetime and the structure of the
singularity is far more complicated than revealed by only radial motions.
Of course, the geometry of the region with $\lambda>(\sin\beta)^{2/3}$
is similar to that of the non-rotating case.

\section{Distribution of angular momentum}

Because the horizon of a supersymmetric $D=5$ black holes is non-rotating, any
angular momentum must be stored in the Maxwell field. Let $\Sigma$ be the same
spacelike hypersurface as before with boundaries at spatial infinity and on the
horizon. An application of Gauss' law to the formula (\ref{angular}) for
the total angular momentum associated with the Killing vector field
$m=2\partial_\psi$ yields
\be\label{dissone}
J= {1\over 4\pi G_D}\int_\Sigma dS_m R^m{}_n m^n + J_H
\ee
where
\be\label{jones}
J_H = {1\over 16\pi G_5} \oint_H dS_{mn}D^m m^n\, .
\ee
It would be natural to associate the surface integral $J_H$ with the
angular momentum due to rotation of the horizon, in which case it would have to
vanish for a non-rotating horizon. However, if there is any contribution to the
bulk integral of (\ref{dissone}) just outside the horizon (and there is) the
existence of the isometry of the near-horizon solution that exchanges the
interior and exterior regions implies that there must be angular momentum in the
Maxwell field {\sl inside} the horizon. This being the case it would be
surprising if $J_H$ did vanish. In fact, it does not, as  shall now verify 
by a direct calculation using the near-horizon metric (\ref{nearhor}). 
Note first that in these coordinates
\be
k= \mu^{-1/2}\partial_u \qquad m= 2\partial_{\psi'}\, .
\ee
 
We begin with the formula
\be\label{meazur}
dS_{mn} = d{\cal A}(k_m n_n - k_nn_m)
\ee
where $n$ is any null vector field with $k\cdot n=-1$ on the horizon. A suitable
choice is
\be
n = \mu^{-1/2}\cos\beta\,\partial_\rho\, .
\ee
Because $k$ is Killing and commutes with $m$ we have
\be
(k_m n_n - k_nn_m)D^mm^n = -n\cdot\partial(k\cdot m) = 2\cos\beta\sin\beta
\ee
and hence
\be
J_H = -{1\over 8\pi G_5} \cos\beta \sin\beta\, {\cal A}\, .
\ee
Now the 3-volume of the squashed 3-sphere horizon is
\be
{\cal A} = 2\pi^2 \mu^{3/2} \cos\beta\, ,
\ee
so, using $j=\sin\beta \mu^{3/2}$ and the formula (\ref{angmom}) for the total
angular momentum $J$ we see that
\be
J_H = -{1\over2}\cos^2\beta\, J
\ee
Not only is this non-zero, it is also a {\sl negative} fraction of the total
angular momentum.

As a check on this result we shall now calculate the bulk contribution
on the spacelike hypersurface $\Sigma$,
\be
J_\Sigma = {1\over 4\pi G_D}\int_\Sigma dS_m R^m{}_n m^n\, .
\ee
To do so we observe that the value of $J_H$ is independent of the value of $u$
at which $\Sigma$ meets the horizon, because $\partial_u$ is Killing. We may
therefore drag the boundary $H$ of $\Sigma$ back to $u=-\infty$ along the orbits
of $\partial_u$ on the horizon, which are its null geodesic generators, without
affecting the value of $J_H$. In this limit the surface $\Sigma$ can be chosen
to be the hypersurface of constant $t$ in the metric (\ref{fivemet}).
Using the Einstein equation we then have
\be
J_\Sigma = {1\over2G_5}\int d^4x \sqrt{-g}g^{0p}e_p{}^a e_\psi{}^c
F_{ab}F^b{}_c
\ee 
{}From the explicit form of the metric (\ref{fivemet}), and choosing the basis
one-forms to be
\be
e^0 = H^{-1}\left(dt + {j\sigma_3\over 2r^2}\right)\, ,
\qquad e^r = H^{1/2}dr\, ,\qquad e^I = {1\over2}rH^{1/2} \sigma_I \qquad 
(I=1,2,3),
\ee
we find that
\be
J_\Sigma = {1\over2G_5}\int d^4x \left[{H^2\sin\theta\over
8r^3}\left(r^6-j^2H^{-3}\right)F_{r0}\left(F_{r0} e_\psi{}^0 + F_{r3}
e_\psi{}^3\right) + {j\sin\theta\over 4r}\left(e_\psi{}^0 F_{r0} + e_\psi{}^3
F_{r3}\right)^2\right]
\ee
Using
\be
F= \sqrt{3} e^r\wedge \left[e^0 {\mu\over r^3} H^{-3/2} -e^3{j\over
r^3}H^{-2}\right] + \sqrt{3} e^1\wedge e^2 \left[ {j\over r^2}H^{-2}\right] 
\ee
we compute
\bea
F_{r0}\left(F_{r0} e_\psi{}^0 + F_{r3} e_\psi{}^3\right) &=&
{3j\mu H^{-4}\over 2r^6}\nn
\left(e_\psi{}^0 F_{r0} + e_\psi{}^3 F_{r3}\right)^2 &=& 3j^2H^{-5}\over 4r^6
\eea
and hence 
\be
J_\Sigma = {1\over2G_5}\int d^4x\, {3\over 16}\sin\theta 
\left[ {j^3\over r^7H^4} -{j\mu\over r^3H^2}\right]
\ee
The integrals are now easily done. Using $\sin\beta =
j/\mu^{3/2}$, we find the result
\bea
J_\Sigma = {\pi j\over 4G_5}\left(\sin^2\beta -3\right) =
 \left(1 + {1\over2}\cos^2\beta\right)  J
\eea
This is larger than the total angular momentum $J$, but is precisely such that
$J_\Sigma + J_H =J$, as required. 

The fact that $J_H$ vanishes as $\cos\beta \rightarrow 0$ suggests that the
angular momentum behind the horizon, on a surface of constant $t$ in the metric
(\ref{insidemet}), is distributed in the region between the horizon and the
intersection of the constant $t$ hypersurface with the hypersurface
$\lambda^3=\sin^2\beta$, since this region vanishes in the same limit. We have
not verified this, but since the hypersurface of constant $t$ becomes timelike
for $\lambda^3<\sin^2\beta$ it is not clear how it would make sense to extend 
the bulk integration into this region.

\section{Near-horizon supersymmetry}

The isometry group of the general stationary black hole solution of D=5
supergravity is $\bR\times U(1)\times U(1)$, corresponding to time
translations and rotational invariance within two independent
2-planes. Black holes for which the only non-zero angular momentum
is that associated to the vector field $m=2\partial_\psi$ have a larger
$\bR\times SU(2)\times U(1)$ isometry group. This includes the
supersymmetric rotating black holes, but in this case the time
translations are enhanced to a supergroup generated by the hamiltonian
and four supercharges. This corresponds to a 1/2 breaking of the
supersymmetry of the Minkowski vacuum solution, but it is known that
there is a full restoration of supersymmetry near the horizon
\cite{Renata}. Our aim here is to determine the isometry supergroup of
this near-horizon solution, which is obtained from the full solution
(\ref{fivemet}) by dropping the constant term from $H$. For
convenience we set $\mu=1$ in this
section to get the near-horizon metric
\be\label{nhsone}
ds^2 = -\left(r^2dt^2+{j\over2}\sigma_3\right)^2 + 
{dr^2\over r^2}  + d\Omega_3^2 \, .
\ee
We have already shown that the singularity at $r=0$ is a coordinate singularity
at a Killing horizon of $k=\partial_t$, and that the metric can be continued
through it. However, the  metric (\ref{nhsone}) will suffice present
purposes. Our first task will be to find the Killing spinors admitted by
the near-horizon solution. These are solutions of (\ref{kilspin}). 

The near-horizon solution (\ref{nhsone}) has the same form as the full solution
(\ref{fivesol}) but with
\bea
e^0 &=& r^2dt + {j\over2}\sigma_3 \nn
e^r &=& r^{-1} dr\nn
e^I &=& {1\over2} \sigma_I \qquad (I=1,2,3)
\eea
The Maxwell 2-form $F$ continues to be given by $(\sqrt{3}/2)de^0$. We compute
that
\bea
{1\over4}\omega_{ab}\Gamma^{ab} &=& 
e^0\left(\Gamma^{r0} + {j\over2}\Gamma^{12} - {j\over2} \Gamma^{r3}\right)
+ e^r\left( -  {j\over2} \Gamma^{03}\right) + 
e^1\left({1\over2} \Gamma^{23} + {j\over2} \Gamma^{02}\right)\nn
&& +e^2\left({1\over2} \Gamma^{31} - {j\over2}\Gamma^{01}\right)
+ e^3\left({1\over 2}\Gamma^{12} - {j\over2}\Gamma^{r0}\right)
\eea
and that
\bea
{1\over 4\sqrt{3}}\left(e^a\Gamma^{bc}{}_aF_{bc} - 4e^a\Gamma^bF_{ab}\right)
&&= e^0\left(\Gamma^r + {j\over2}\Gamma^{0r3} -{j\over2}\Gamma^{012}\right)
+ e^r\left( - \Gamma^0 + j\Gamma^3 + {j\over2}\Gamma^{r12}\right)\nn
&&+ e^1\left({1\over2}\Gamma^{1r0} - {j\over2} \Gamma^{1r3}-j\Gamma^2\right)
+ e^2\left({1\over2}\Gamma^{2r0} - {j\over2} \Gamma^{2r3}+ j\Gamma^1\right)\nn
&&+ e^3\left({1\over2}\Gamma^{3r0} - j\Gamma^r  +{j\over2}\Gamma^{123}\right)
\eea
{}From these results we obtain the following Killing spinor equations:
\bea
0 &=& \big[\partial_t + r^2(\Gamma^{r0} + i\Gamma^r)\big] \zeta \nn
0 &=& \big[\partial_r - r^{-1}(i\Gamma^0 +
j\Gamma^{03}-ij\Gamma^3)\big]
\zeta
\nn 0 &=& \big[\partial_\theta - {1\over2}\sin\psi \hat M_1  
+{1\over2}\cos\psi \hat M_2\big]\zeta \nn
0 &=& \big[\partial_\phi + {1\over2}\cos\psi\sin\theta \hat M_1 +
{1\over2}\sin\psi\sin\theta \hat M_2  +  {1\over2}\Gamma^{12} \cos\theta
\big]\zeta\nn 
0 &=& \big[\partial_\psi + {1\over2}\Gamma^{12}  \big]\zeta
\nn
\eea
where
\be
\hat M_1 = \Gamma^{23} + j\Gamma^{02} - i j\Gamma^2\, , \qquad
\hat M_2 = \Gamma^{31} -j\Gamma^{01} + ij\Gamma^1
\ee

All integrability conditions are satisfied, in agreement with \cite{Renata}.
The Killing spinors are
\bea
\zeta^+ &=& r\Omega\eta^+ \nn
\zeta^- &=& \bigg[{1\over r}(1-j\Gamma^{03})-2rt\Gamma^{r0}\bigg]\Omega\eta^-
\eea
where
\be
\Omega = e^{{1\over2}\Gamma^{21}\psi}e^{{1\over2}\Gamma^{13}\theta}
e^{{1\over2}\Gamma^{21}\phi}
\ee
and $\eta^\pm$ are constant spinors satisfying\footnote{Note that $\zeta^-$ is
not an eigenspinor of $i\Gamma^0$.}
\be
i\Gamma^0\eta^\pm = \pm\eta^\pm\, .
\ee

A modification of an argument in \cite{GMT} shows that if $\zeta$ and
$\zeta'$ are Killing spinors then the vector field
\be
v=\bar\zeta\Gamma^a\zeta' \tilde e_a
\ee
is Killing, where $\tilde e_a$ ($a=0,1,2,3,r$) are the basis vector
fields dual to $e^a$. The Killing vector fields found this way are linear
combinations of the vector fields that generate the bosonic subgroup
of the isometry supergroup. Moreover, as shown in \cite{GMT}, the linear
combination associated with any pair of Killing spinors is the same as the
linear combination of bosonic charges appearing in the anticommutator of spinor
charges. The isometry supergroup can therefore be determined (modulo purely
bosonic factors) from a knowledge of the Killing spinors.

To proceed, we note that
\be
\tilde e^a = (r^{-2}\partial_t,\quad r\partial_r,\quad 2\xi_1^R, \quad
2\xi_2^R,\quad 2\xi_3^R- jr^{-2}\partial_t)
\ee
where
\bea
\xi_1^R &=& -\sin\psi\partial_\theta + \cos\psi\cosec\theta\partial_\phi
-\cot\theta\cos\psi\partial_\psi \nn
\xi_2^R&=& \cos\psi\partial_\theta + \sin\psi\cosec\theta\partial_\phi 
-\cot\theta\sin\psi\partial_\psi \nn
\xi_3^R &=& \partial_\psi \, .
\eea
These are the three left-invariant vector fields $\xi_I^R$ $(I=1,2,3)$
generating right translations on $SU(2)$; they are dual to the basis of
left-invariant 1-forms $\sigma_I$ in the sense
that $(\xi_I^R,\sigma_J)=\delta_{IJ}$. We will also need the right-invariant
vector fields
\bea
\xi^L_1 &=& \sin\phi\partial_\theta + \cot\theta\cos\phi \partial_\phi 
-\cos\phi\cosec\theta \partial_\psi\nn
\xi^L_2 &=& \cos\phi\partial_\theta - \cot\theta\sin\phi\partial_\phi 
+ \sin\phi\cosec\theta \partial_\psi\nn
\xi^L_3 &=& \partial_\phi\, ,
\eea
which commute with $\xi_I^R$ and generate left translations on $SU(2)$. The
non-vanishing commutation relations among all six vector fields are
\be\label{comrel}
{[\xi_i^R,{\xi_{j}^{R}}]}=-\epsilon_{ijk}\xi_k^R\, , \qquad
{[{\xi_{i}^{L},\xi_{j}^{L}}]}=\epsilon_{ijk}\xi_k^L
\ee

Now, we know that the following vectors fields are Killing:
\bea\label{killhere}
v^{++} &=& (\bar\zeta^+ \Gamma^a \zeta^+) \tilde e_a \nn
v^{+-} &=& (\bar\zeta^+ \Gamma^a \zeta^-) \tilde e_a \nn
v^{--} &=& (\bar\zeta^- \Gamma^a \zeta^-) \tilde e_a\, .
\eea
A computation yields the result 
\bea\label{vees}
v^{++} &=& (\bar\eta_+\Gamma^0\eta_+)\, r^2 \tilde e_0 \nn
v^{+-} &=& (\bar\eta_+\Gamma^r\eta_-)(\tilde e_r -2r^2t\,\tilde e_0) 
+ (\bar\eta_- \Omega^{-1}\Gamma^1\Omega \eta_-) \tilde e_1 \nn
&& + (\bar\eta_- \Omega^{-1}\Gamma^2\Omega \eta_-) \tilde e_2
+ (\bar\eta_- \Omega^{-1}\Gamma^3\Omega \eta_-) (\tilde e_3 + j\tilde e_0)\nn
v^{--} &=& (\bar\eta_-\Gamma^0\eta_-)\big[ r^{-2}(1+j^2 + 4r^4 t^2)\tilde e_0
-4t\tilde e_r + 2jr^{-2} \tilde e_3 \big]\, .
\eea
But
\be
\Omega^{-1}\Gamma^i\Omega \equiv  \Gamma^j {R_j}^i(\Omega)
\ee
where $R(\Omega)$ is a matrix such that
\be\label{lem2} 
{R_j}^i(\Omega)\xi_i^R=\xi_j^L\, .
\ee
It follows that
\bea
v_{++} &=& (\bar\eta_+\Gamma^0\eta_+)\,k \nn
v_{+-} &=& (\bar\eta_+\Gamma^r\eta_-)\ell + 2(\bar\eta_-\Gamma^1\eta_-)\xi^L_1
+ 2(\bar\eta_-\Gamma^2\eta_-)\xi^L_2 + 2(\bar\eta_-\Gamma^3\eta_-)\xi^L_3 \nn
v_{--} &=& (\bar\eta_-\Gamma^0\eta_-) m 
\eea
where
\bea
k &=& \partial_t \nn
\ell &=& r\partial_r -2t\partial_t \nn
m &=& r^{-4}(1-j^2)\partial_t + 4t^2\partial_t -4tr\partial_r +
4jr^{-2}\partial_\psi
\eea
We thus deduce that the vector fields $\xi^L$ are Killing, as must be
$k,\ell,m$. The latter commute with $\xi^L$, and also with $\xi^R$. Their
commutation relations with each other are
\be
[\ell,k] = 2k \qquad [\ell,m] = -2m \qquad [k,m] =-4\ell
\ee
which are those of $sl(2;\bR)$. 
We conclude that the anticommutator of supersymmetry charges closes on a set of
bosonic charges that generate $Sl(2;\bR)\times SU(2)$. The vector field 
$\xi_3^R$ is also Killing, as are $\xi_1^R$ and $\xi_2^R$ when $J=0$, but these
vector fields are {\sl not} constructible from Killing spinors and so are not
associated with charges in the supersymmetry algebra. This explains is why it is
possible to break rotational symmetry without losing supersymmetry. 

{}From the above information we deduce that the isometry supergroup of the
near-horizon limit of a $D=5$ supersymmetric black hole is $SU(1,1|2)\times
SU(2)$ for $J=0$ and $SU(1,1|2)\times U(1)$ for $J\ne 0$. 

\section{Discussion}

The bosonic equations of $D=5$ supergravity can be viewed as the special $\nu=1$
case of the general $D=5$ Einstein-Maxwell-CS equations derived from the
Lagrangian
\be\label{cremnu}
L= {1\over 16\pi G_5} \left[\sqrt{-g}(R -F^2) -
{2\nu\over 3\sqrt{3}}\varepsilon^{mnpqr}A_mF_{np}F_{qr}\right]
\ee
where $\nu$ is a CS coupling constant. For $\nu=1$ there is a class of
supersymmetric black hole solutions parameterized by the charge $Q$ and an
angular momentum $J$, with
the mass $M$ being fixed in terms of the charge as a
consequence of the fact that it must saturate the bound (\ref{bound}). Here we
have uncovered a number of surprising features of these solutions. The horizon 
is non-rotating even when the angular momentum is non-zero. This is possible
because angular momentum can be stored in the Maxwell field, but another
surprise is that a negative fraction of the total is stored behind the horizon. 
Physically, one might think that these two results are related. The fact
that the horizon is non-rotating is a statement that the null generators
of the horizon experience no angular frame dragging with respect to infinity.
One might be able to regard this as the result of a cancellation of
the individual dragging effects which the angular momentum distributions
inside and outside of the horizon would have produced.

The derivation in \cite{GKLTT} of the bound (\ref{bound}) on the mass of black
holes of $D=5$ supergravity made crucial use of the fact that $\nu=1$. However,
the (non-rotating) Tangherlini black hole is a solution of the
Einstein-Maxwell-CS equations for any value of $\nu$, and its extremal
mass is still given by the formula $M=(\sqrt{3}/2)|Q|$. It also has vanishing
surface gravity, and so cannot decrease its mass by Hawking radiation. It might
therefore appear that the bound $M\ge (\sqrt{3}/2)|Q|$ must be valid for {\sl
any} value of $\nu$. It is possible that the methods of \cite{sparling} may
serve to derive it for $\nu<1$, but there are some reasons to think that
it may fail when $\nu>1$. The way in which the proof in \cite{GKLTT} 
fails when $\nu\ne1$ shows that any violation must involve field configurations
for which $F\wedge F$ is non-zero. This condition can be  satisfied for
perturbations which redistribute the angular momentum in the Maxwell field,
suggesting that any instability of the (spherically symmetric) Tangherlini black
hole may involve rotation. It might seem unlikely that a black hole solution of
zero Hawking temperature and spherical symmetry could be unstable, but a $\nu=1$
Tangherlini black hole is only marginally stable because there is a rotating
black hole with precisely the same total mass. 

As far we are aware, the analogue of the Kerr-Newman solution for
$D=5$ Einstein-Maxwell ($\nu=0$) is not yet known, but it is likely to
to be a straightforward extension of the $D=5$ analogue of the Kerr solution
found in \cite{MP}. If so, the mass of black hole solutions with vanishing
surface gravity can be expected to be a strictly increasing function of either
of their two angular momenta. Given the known solutions for $\nu=1$ it then
seems likely that the increase in the energy with at least one angular momentum
$J$ decreases as $\nu$ increases from zero to $1$, so as to become independent 
of $J$ at $\nu=1$. If this is the case then an extrapolation to $\nu>1$ might
lead to an energy that decreases with increasing $J$, creating an instability
of the Tangherlini black hole against a perturbation in which photons carry away
both energy and angular momentum to infinity.

It is interesting to frame this discussion in terms of the first law,
which we derived in section 2. As we are discussing extremal black holes
($\kappa=0$) and processes which leave the charge unchanged ($dQ=0$) the first
law reduces to $dM= {\bf \Omega}_H\cdot d{\bf J}$. As long as the angular
velocities of the horizon are nonvanishing and positive (i.e., the rotation
is in the same sense as the angular momentum) when $\bf J$ are
nonvanishing, a further increase of the angular momenta produces an increase in
the mass. This corresponds to the situation which we expect to hold for $\nu=0$,
and in fact $\nu<1$. We know that at precisely $\nu=1$, the angular velocity of
the horizon can vanish even though the  angular momentum is nonvanishing. This
produces the marginal situation where variations of the angular momentum 
leave the mass unchanged. The instability that is speculated to arise
for $\nu>1$ would require that at least one of the ${\bf \Omega}_H$ is {\it
negative} when $\nu>1$. That is, the horizon would be rotating in the opposite
sense to the angular momentum. In this case, a further increase of the angular
momentum could be used to reduce the black hole mass.

Pursuing the speculation on frame dragging above, we would have the
following physical picture. Imagine we start with the $\nu=0$ solution
with, for simplicity, only the angular momentum \reef{angmom} associated with
$2\partial_\psi$ nonvanishing. We expect that both the corresponding angular
velocity $\Omega_H$ and the interior angular momentum, measured by $J_H$, will 
be nonvanishing and positive. As the CS coupling is increased, the solution will
change, presumably such that both $J_H$ and $\Omega_H$ are reduced.
For some $\nu<1$, $J_H$ would reach zero and become negative, but with $\Omega_H$
still positive. The latter would only reach zero at $\nu=1$. A further increase
in the CS coefficient to $\nu>1$ would then, according to our conjecture, 
produce an unstable solution for which both $J_H$ {\sl and} $\Omega_H$ are
negative. The rearrangement in the distribution of the angular momentum would
then be the essential difference in the solutions for different values of the
CS coupling.

The instability conjectured to exist here for a certain range of
the five-dimensional Chern-Simons coupling is in some respects reminiscent
of the instability in four-dimensional theories
with massive charged vectors \cite{alex}. There a magnetically charged
Reissner-Nordstrom black hole becomes unstable to developing massive vector hair
which is not spherically symmetric when the vector mass reaches a critical
value\footnote{In this case, the critical vector mass is correlated
to the black hole size.}. While these instabilities appear to violate
the `standard' no-hair theorems, the latter are evaded by the `nonstandard'
matter content of the theories in question \cite{whoo}. Of course,
the instability in the present case could be ruled out by an extension of the
uniqueness theorems to $D=5$, but there is no evidence that these theorems apply
in $D=5$ for arbitrary $\nu$. One is free to surmise that uniqueness also
applies only for $\nu\le1$. It would not be surprising to learn once again that
supersymmetry is associated with a borderline between stability and instability.

\vskip 0.5cm
\noindent
{\bf Acknowledgements}: This work was begun during the workshop
``Dualities in String Theory'' hosted by the ITP of the University of California
at Santa Barbara. The authors thank the ITP for hospitality. They also thank
Fay Dowker and Gary Gibbons for helpful conversations. JPG thanks the
EPSRC for financial support. RCM was supported by NSERC of Canada
and Fonds FCAR du Qu\'ebec. Research at the ITP was supported by
NSF Grant PHY94-07194.
 


\bigskip
\begin{thebibliography}{99}

\bibitem{Tang}
{}F.R. Tangherlini, {\sl Schwarzschild field in $n$ dimensions and the
dimensionality of space problem}, Nuovo Cim. {\bf 27} (1963) 636. 

\bibitem{MP}
R.C. Myers and M.J. Perry, {\sl Black holes in higher dimensional space-times},
Ann. Phys. (NY) {\bf 172} (1986) 304.

\bibitem{BCH}
J. Bardeen, B. Carter and S.W. Hawking, {\sl The four laws of black hole
mechanics}, Commun. Math. Phys. {\bf 31} (1973) 161.

\bibitem{Wald}
R.M. Wald, {\sl Black hole entropy is Noether charge}, Phys. Rev. {\bf D48}
(1993) 3427.

\bibitem{GH}
G.W. Gibbons and C.M. Hull, {\sl A Bogomolnyi bound for General Relativity
and solitons in N=2 supergravity}, Phys. Lett. {\bf 109B} (1982) 190.

\bibitem{WO}
E. Witten and D. Olive, {\sl Supersymmetry algebras that include topological
charges}, Phys. Lett. {\bf 78B} (1978) 97.

\bibitem{Tada}
J.P. Gauntlett and T. Tada, {\sl Entropy of four-dimensional rotating BPS
dyons}, Phys. Rev. {\bf D55} (1997) 1707.

\bibitem{Waldb}
R.M. Wald, {\sl The first law of black hole mechanics}, gr-qc/9305022.

\bibitem{Cremmer}
E. Cremmer, {\sl Supergravities in 5 dimensions}, in {\it Superspace and
Supergravity}, eds. S.W. Hawking and M. Ro{\v c}ek (CUP 1980).

\bibitem{GKLTT}
G.W. Gibbons, D. Kastor, L.A.J. London, P.K. Townsend and J. Traschen, {\sl
Supersymmetric self-gravitating solitons}, Nucl. Phys. {\bf B416} (1994) 850.

\bibitem{lessix}
J.C. Breckenridge, D.A. Lowe, R.C. Myers, A.W. Peet, A. Strominger and 
C. Vafa, {\sl Macroscopic and microscopic entropy of near-extremal spinning
black holes}, Phys. Lett. {\bf 381B} (1996) 423.

\bibitem{tsey}
A.A. Tseytlin, {\sl Extreme dyonic black holes in string theory}, Mod. Phys.
Lett. {\bf A11} (1996) 689.

\bibitem{myers}
J.C. Breckenridge, R.C. Myers, A.W. Peet and C. Vafa, {\sl D-branes and
spinning black holes}, Phys. Lett. {\bf 391B} (1997) 93. 

\bibitem{sabra}
A.H. Chamseddine and W.A. Sabra, {\sl Metrics admitting Killing spinors in five
dimensions}, Phys. Lett. {\bf 426B} (1998) 36.

\bibitem{HSen}
G. Horowitz and A. Sen, {\sl Rotating black holes which saturate a
Bogomolnyi bound}, Phys. Rev. {\bf D53} (1996) 808.

\bibitem{Carter}
B. Carter, in {\it Black Holes}, eds. C. and B. DeWitt, Gordon and
Breach, New York, 1973 (1972 Les Houches Summer School Lectures)

\bibitem{Gibbons}
G.W. Gibbons, {\sl Aspects of supergravity theories}, in {\it Supersymmetry,
Supergravity and Related Topics}, eds. F. Del Aguila, J.A. de Azc{\'a}rraga and
L.E. Iba{\~n}ez, World Scientific, Singapore, 1985.

\bibitem{Kal}
P. Claus, M. Derix, R. Kallosh, J. Kumar, P.K. Townsend and A. Van Proeyen, {\sl
Superconformal mechanics and black holes}, hep-th/9804177.

\bibitem{Ferrara}
A. Chamseddine, S. Ferrara, G.W. Gibbons and R. Kallosh, {\sl Enhancement of
supersymmetry near the 5D black hole horizon}, Phys. Rev. {\bf D55} (1977)
3647.   

\bibitem{Renata}
R. Kallosh, A. Rajaraman and W.K. Wong, {\sl Supersymmetric rotating black 
holes and attractors}, Phys. Rev. {\bf D55} (1997) 3246.

\bibitem{GMT}
J.P. Gauntlett, R.C. Myers and P.K. Townsend, {\sl Supersymmetry of rotating
branes}, hep-th/9809065, to appear in Physical Review D.

\bibitem{Hawking}
S.W. Hawking, {\sl Black Holes in General Relativity}, Commun. Math. Phys. {\bf
25} (1972) 152.

\bibitem{GKK}
G.W. Gibbons, R. Kallosh and B. Kol, {\sl Moduli, scalar charges, and the first
law of black hole thermodynamics}, Phys. Rev. Lett. {\bf 77} (1996) 4992.

\bibitem{Carterctc}
B. Carter, {\sl Global structure of the Kerr family of gravitational fields},
Phys. Rev. {\bf D174} (1968) 1559; {\sl Domains of Stationary Communications in
Spacetime}, Gen. Rel. Grav. {\bf 9} (1978) 437.

\bibitem{sparling}
O.M. Moreschi and G.A.J. Sparling, {\sl On the positive energy theorem
involving mass and electromagnetic charges}, Commun. Math. Phys. {\bf 95}
(1984) 113.

\bibitem{alex}
S.A. Ridgway and E.J. Weinberg, {\sl Are All Static Black Hole Solutions
Spherically Symmetric?}, Gen. Rel. Grav. {\bf 27} (1995) 1017 [gr-qc/9504003];
{\sl Static Black Hole Solutions Without Rotational Symmetry},
Phys. Rev. {\bf D52} (1995) 3440 [gr-qc/9503035].

\bibitem{whoo}
D. Kastor and J. Traschen, {\sl Horizons Inside Classical Lumps},
Phys. Rev. {\bf D46} (1992) 5399 [hep-th/9207070].


\end{thebibliography}
\end{document}